\documentclass[
twocolumn,
aps,prd,
superscriptaddress,
groupedaddress,
floatfix,
longbibliography]{revtex4-2}

\usepackage[utf8]{inputenc}
\usepackage{hyperref}
\usepackage{dsfont}
\usepackage{graphicx}% Include figure files
\usepackage[caption=false]{subfig}
\usepackage{epstopdf}
\usepackage[dvipsnames]{xcolor}
\usepackage{amsmath,amssymb}

\definecolor{C1}{RGB}{0,206,206}
\definecolor{C2}{RGB}{251, 77, 61}
\definecolor{C3}{HTML}{5AA9E6}
\definecolor{C4}{RGB}{202, 21, 81}

\hypersetup{colorlinks=true, linkcolor=C2, citecolor=C2, urlcolor=C2}

\newcommand*{\f}{\frac}
\newcommand*{\mc}{\mathcal}
\newcommand*{\dg}{\dagger}
\newcommand*{\mf}{\mathfrak}
\newcommand*{\mbb}{\mathds}

\newcommand*{\design}{\mathsf{t}}
\newcommand*{\haar}{\mathsf{h}}
\DeclareMathOperator{\tr}{tr}
\DeclareMathOperator{\swap}{\text{\scshape{swap}}}
\DeclareMathOperator{\FS}{\mf{S}}
\newcommand*{\markov}{{\scriptscriptstyle{(\mathrm{M})}}}

\usepackage{fdsymbol}
\newcommand*{\bdiamond}{\smallblackdiamond}

\usepackage{amsthm}
\newtheorem*{theorem*}{Theorem}
\newtheorem{theorem}{Theorem}
\newtheorem{lemma}[theorem]{Lemma}

\usepackage{tikz}
\usetikzlibrary{fadings,decorations.pathreplacing}

\begin{document}
\title[]{Markovianization with approximate unitary designs}

\author{Pedro Figueroa--Romero}
\email[]{pedro.figueroaromero@monash.edu}
\author{Felix A. Pollock}
\author{Kavan Modi}
\email[]{kavan.modi@monash.edu}
\affiliation{School of Physics \& Astronomy, Monash University, Clayton, Victoria 3800, Australia}
\date{\today}

\begin{abstract}
\vspace{1em}\section*{\abstractname}\vspace{-1em}
Memoryless processes are ubiquitous in nature, in contrast with the mathematics of open systems theory, which states that non-Markovian processes should be the norm. This discrepancy is usually addressed by subjectively making the environment forgetful. Here we prove that there are physical non-Markovian processes that with high probability look highly Markovian for all orders of correlations; we call this phenomenon Markovianization. Formally, we show that when a quantum process has dynamics given by an approximate unitary design, a large deviation bound on the size of non-Markovian memory is implied. We exemplify our result employing an efficient construction of an approximate unitary circuit design using two-qubit interactions only, showing how seemingly simple systems can speedily become forgetful. Conversely, since the process is closed, it should be possible to detect the underlying non-Markovian effects. However, for these processes, observing non-Markovian signatures would require highly entangling resources and hence be a difficult task.
\end{abstract}

\maketitle

\setlength{\parskip}{0.5em}

\section{Introduction}
A foundational question of modern physics is to understand the origins of irreversibility~\cite{Gogolin_2016}. In particular, to determine whether fundamental laws, which are fully reversible, are consistent with phenomena like equilibration and thermalization. The dynamical version of this conundrum concerns the emergence of forgetful processes from isolated ones. In quantum mechanics, an isolated process is unitary, and cannot lose information; past behaviour in one part of the system will always be remembered, eventually returning to influence the future.

However, there are many ways in which nature manifests forgetful processes, where a system's evolution is determined with a seeming disregard to its previous interactions with its surroundings. For example, a carbon atom does not typically remember its past and behaves like any other carbon atom. Such processes are not isolated, and the general intuition is that the dynamics of a system, in contact with a large environment, can be approximately described as memoryless~\cite{breuer2002theory}. Yet, formal derivations of memory-less quantum processes require several assumptions about the coupling strength with the environment, the timescales of dynamical correlations, and an infinite-dimensional reservoir. For finite-sized environments, this can only be achieved exactly by continually refreshing (discarding and replacing) the environment’s state, i.e., artificially throwing away information from the environment. The problem this poses is akin to the one made by the Fundamental Postulate of Statistical Mechanics~\cite{Gogolin_2016}, which a-priori sets the probabilities of a closed system to be in any of its accessible microstates as equal.

Thus the foundational question remains open: can forgetful processes arise from isolated processes without any artificial discarding of information? Because forgetful processes are often called Markovian, we refer to the mechanism for forgetting as \textit{Markovianization}, in the same spirit as the terms equilibration and thermalization~\cite{Popescu2006, PhysRevE.79.061103, gemmer2009, Vinayak_2012, PhysRevE.86.031101, Masanes2013, Gogolin_2016}. Indeed, Markovianization is likely to come about through mechanisms intimately related to these other processes. For instance, dissipative Markov processes have fixed points to which the system relaxes; this is a mechanism for equilibration, and also possibly for thermalization. We have previously argued for the emergence of Markovianization for mathematically typical processes, using averages with respect to the Haar measure~\cite{FigueroaRomero2019almostmarkovian}; however, such processes are far from physically typical~\cite{Gogolin_2016}.

In this paper, we identify a class of isolated physical processes which approximately Markovianize in a strong sense, where even the multi-time quantum correlations vanish. To do so, we employ large deviation bounds for approximate unitary designs derived by R. Low~\cite{Low_2009}, and apply them to the process tensor formalism~\cite{PhysRevA.97.012127, Simon_Operational, Quolmogorov}, which describes quantum stochastic processes. We show that, similar to the way in which quantum states thermalize, quantum processes can Markovianize in the sense that they can converge to a class of typical processes, satisfying a meaningful large deviation principle whenever they are undergone within a large environment and under complex enough ---but not necessarily fully random--- dynamics. As a proof of principle, we employ a recent efficient construction of approximate unitary designs with quantum circuits~\cite{Winter_HamDesign} to illustrate how a dilute gas would quickly Markovianize. These results directly impose bounds on complexity and timescales for standard master equations employed in the theory of open systems. Finally, we discuss possible extensions of our results to many-body systems with time-independent Hamiltonians. Our results are timely given the ever-increasing interest and relevance in determining the breakdown of the Markovian approximation in modern experiments~\cite{Gessner2014, PhysRevLett.114.090402, Josh_2019, Winick_2019}.

\section{Results}
\subsection{Quantum stochastic processes}
A classical stochastic process on a discrete set of times is the joint probability distribution of a time-ordered random variable, $\mbb{P}(x_k,\dots,x_0)$. A process is said to have finite memory whenever the state of the system at a given time is only conditionally dependent on its previous $m$ states: $\mbb{P}(x_k|x_{k-1}, \dots,x_0) =\mbb{P}(x_k|x_{k-1}, \dots, x_{k-m})$. Here, $m$ is the Markov order; when $m=1$ the process is called Markovian, and when $m=0$ the process is called random. Finite memory processes, and in particular Markov processes, have garnered significant attention in the sciences for two principal reasons. First, the complexity of a process grows with the Markov order and thus it is easier to work with finite memory processes. Second, many physical processes tend to be well approximated by those with finite memory.

Generalisations of Markov processes and Markov order to the quantum realm have been plagued with technical difficulties~\cite{RevModPhys.88.021002}, which have their origin in the fundamentally invasive nature of quantum measurement. However, recently, a generalized and unambiguous characterization of quantum stochastic processes within the process tensor framework~\cite{PhysRevA.97.012127, PhysRevLett.120.040405} has paved the way to alleviating these difficulties. The success of this framework lies in generalising the notion of time-ordered events in the quantum realm.

\begin{figure}[t]
    \centering
    \includegraphics[width=0.475\textwidth]{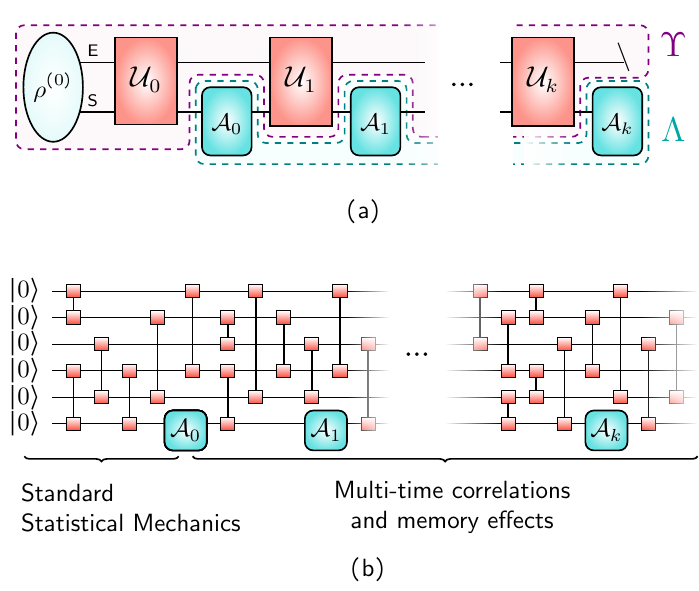}
    \caption{\textbf{Quantum processes and the process tensor.} (a) A $k$-step quantum process $\Upsilon$ on system $\mathsf{S}$ alone is due to the time evolution of an initial system-environment ($\mathsf{SE}$) state $\rho^{(0)}$ with distinct unitary transformations $\mc{U}_i$ with $i=0,1,\ldots,k$. In between each pair of unitaries, an external operation (e.g. a measurement) $\mc{A}_i$ for $i=0,1,\ldots,k$ is applied; this can also be described by a tensor $\Lambda$. (b) An $n$-qubit $\mathsf{SE}$-system ($|0\rangle$ depicting a single qubit) with two-qubit gate interactions (depicted by vertical lines between squares) only: a subsystem qubit is probed at the $i$\textsuperscript{th} step through $\mc{A}_i$. While the standard approach towards typicality or equilibrium properties concerns the whole $\mathsf{SE}$ dynamics and/or a single measurement on system $\mathsf{S}$ as in Standard Statistical Mechanics, we show that complex -- not necessarily uniformly random -- dynamics within large environments will be highly Markovian with high probability.}
    \label{Fig:Processes}
\end{figure}

Consider a system-environment composite $\mathsf{SE}$ of dimension $d_{\mathsf{SE}}=d_\mathsf{S}d_\mathsf{E}$ with an initial state $\rho^{(0)}$ that undergoes a evolution $\mathcal{U}_0$. An intervention $\mc{A}_0$ is then made on the system $\mathsf{S}$ alone, followed by evolution $\mathcal{U}_1$. For concreteness, onward we will consider $\mathcal{U}_i \neq \mathcal{U}_j$. Then a second intervention $\mc{A}_1$ on $\mathsf{S}$ alone. This continues until a final intervention $\mc{A}_k$ is performed following $\mathcal{U}_k$. A quantum event $x_i$ at the $i$\textsuperscript{th} time step corresponds to an outcome of the corresponding intervention, and is represented by a completely positive (\texttt{CP}) map $\mc{A}_{x_i} (\cdot) := \sum_\nu {A}_{x_i}^\nu (\cdot) A_{x_i}^{\nu \dg}$ with Kraus operators $\{A^\nu\}$ satisfying $\sum{A}^{\nu \dg} {A}^\nu \leq \mbb1$. In other words, an intervention is the action of an instrument $\mathcal{J} = \{\mc{A}_{x_i}\}_{x_i}^{X_i}$ where $A_i = \sum_{x_i} \mc{A}_{x_i}$ is a completely positive trace preserving (\texttt{CPTP}) map. This is depicted schematically in Fig.~\ref{Fig:Processes}. In general, the evolution $\mathcal{U}$ is allowed to be a \texttt{CPTP} map on $\mathsf{SE}$. In this paper, however, we are interested in an isolated $\mathsf{SE}$, where the $\mathcal{U}$s are unitary transformations: $\mathcal{U} (\cdot) := U (\cdot) U^{\dag}$, with $U$ a unitary operator.

The probability to observe a sequence of quantum events is given by
\begin{gather}\nonumber
\mbb{P}(x_k,\dots,x_0|\mathcal{J}_k,\dots,\mathcal{J}_0)= \tr\left[\mathcal{A}_{x_k} \mathcal{U}_{k-1} \!\dots \mathcal{A}_{x_0} \mathcal{U}_0 \rho^{(0)}\right].
\end{gather}
This can be rewritten, clearly separating the influence of the environment from that of the interventions, in a multi-time generalization of the Born rule~\cite{Oreshkov2012, Costa_2016, Shrapnel_2018}:
\begin{gather}
\mbb{P}(x_k,\dots,x_0|\mathcal{J}_k,\dots,\mathcal{J}_0)= \tr\left[\Upsilon \Lambda^\mathrm{T}\right],
\label{eq:process}
\end{gather}
where $\text{T}$ denotes transpose, $\Lambda := \mathcal{A}_{x_0} \otimes \dots \otimes \mathcal{A}_{x_k}$, and the effects on the system due to interaction with the environment have been isolated in the so-called process tensor $\Upsilon$. We have depicted $\Upsilon$ and $\Lambda$ in Fig.~\ref{Fig:Processes}(a) as the red and green comb-like regions, respectively. A circuit depiction of the same process $\Upsilon$, along with the instruments $\Lambda$ is given in  Fig.~\ref{Fig:Processes}(b).

Maps like the process tensor are abstract objects with many different representations~\cite{Simon_Operational}. In this manuscript, for convenience, we work with the Choi state representation~\cite{Simon_Operational, watrous_2018} of the process tensor, shown in Eq.~\eqref{eq: Upsilon} of the Methods section. The process tensor $\Upsilon$ is a complete representation of the stochastic quantum process, containing all accessible multi-time correlations~\cite{PhysRevLett.122.140401, PhysRevA.99.042108, Phil_MemStr, arXiv:1811.03722}. Similarly, the tensor $\Lambda$ contains all of the details of the instruments and their outcomes. This tensor, in general, is also a quantum comb, where the bond represents information fed forward through an ancillary system. Finally, the process tensor can be formally shown to be the quantum generalisation of a classical stochastic process, satisfying a generalized extension theorem with consistency conditions for a family of joint probabilities to guarantee the existence of an underlying continuous quantum stochastic process~\cite{Quolmogorov}, and reducing to classical stochastic process in the correct limit~\cite{PhysRevA.100.022120, arXiv:1907.05807}.

\subsection{Measuring non-Markovianity}
The convenience of using the Choi state $\Upsilon$ is that it translates temporal correlations between timesteps into spatial correlations. Furthermore, as detailed in the Methods section on the process tensor, $\Upsilon$ can be efficiently described when written as a matrix product operator~\cite{PhysRevA.97.012127, Verstraete_2004}, whose bond dimension represents the dimension of a quantum environment that could mediate the non-Markovian correlations. In particular, when the bond dimension is one, the process is Markovian. Specifically, a process $\Upsilon^\markov$ is Markovian if and only if it has the form
\begin{equation}
    \Upsilon^\markov=\mc{E}_{1:0}\otimes\cdots\otimes\mc{E}_{k:k-1},
    \label{eq: upsilon markov}
\end{equation}
with $\mc{E}_{j:i}$ a \texttt{CPTP} map on the system connecting the $i$\textsuperscript{th} to the $i+1$\textsuperscript{th} time~\cite{PhysRevLett.120.040405, Simon_Operational}. This quantum Markov condition in Eq.~\eqref{eq: upsilon markov} allows for a precise quantification of memory effects; it is fully consistent with the classical Markov condition, and contains all of the popular witnesses of quantum non-Markovianity~\cite{RevModPhys.88.021002}. Importantly, it allows for operationally meaningful measures of non-Markovianity: for instance, the relative entropy of the process tensor with respect to its marginals, which happen to be the closest Markovian process tensor, i.e. $\mc{N}_\mc{S} :=\min_{\Upsilon^\markov}\mc{S}(\Upsilon\|\Upsilon^\markov)$, quantifies the probability of mistaking $\Upsilon$ and $\Upsilon^\markov$, which decreases in the number of realisations of the process $n$ as $\exp(-n\mc{N}_\mc{S})$.

For the current considerations, a natural choice is the so-called diamond norm. Just as trace distance is a natural metric for differentiating two quantum states, in the sense of having a clear operational definition, the natural distance for differentiating two quantum channels is the diamond norm, which allows for the use of additional ancillas~\cite{PhysRevA.71.062310}. We are interested in optimally differentiating between a non-Markovian process from a Markovian one, which leads to the multi-time diamond distance:
\begin{gather}
 \mc{N}_\bdiamond:=\f{1}{2}\min_{\Upsilon^\markov}\|\Upsilon-\Upsilon^\markov\|_\bdiamond,
 \label{Eq: def nM diamond}
\end{gather}
where $\|X\|_\bdiamond:=\sup_{\{\mc{O}_i\},i}\|\sum_i\tr[\mc{O}_i{X}\otimes\mbb1]|i\rangle\!\langle{i}|\|_1$ is a generalized diamond norm~\cite{Phil_MemStr, PhysRevA.80.022339}, with the supremum over $i\geq1$ and a set of CP maps $\{\mc{O}_i\}$. This definition generalizes the diamond norm for quantum channel distinguishability~\cite{Kitaev} (also called cb-norm~\cite{Paulsen_diamond} or completely bounded trace norm~\cite{watrous_2018}), reducing to it for a single step process tensor, and similarly being interpreted as the optimal probability to discriminate a process from the closest Markovian one in a single shot, given any set of measurements, which can be made together with an ancilla.

Vanishing non-Markovianity in Eq.~\eqref{Eq: def nM diamond} would imply that the process must have the form of Eq.~\eqref{eq: upsilon markov}. The derivations of such processes make ad-hoc assumptions such as artificially refreshing the environment between time-steps (i.e., assumption of an infinite bath) that render approximations such as Born-Markov. Classical processes additionally require randomness injection by hand for stochasticity. Here, we show that a class of underlying quantum mechanisms lead to the emergence of Markovianity without ad-hoc assumptions. Namely, We show that the above measure of non-Markovianity in Eq.~\eqref{Eq: def nM diamond} vanishes as the global $\mathsf{SE}$ dynamics becomes more complex. This is entirely analogous to entanglement being the underlying mechanism explaining the emergence of statistical mechanics from quantum dynamics alone and accounting for the artificial postulate of equal a-priori probabilities~\cite{Popescu2006}.

\subsection{Markovianization with unitary designs}
The generic form of open quantum dynamics is non-Markovian, but, despite this, it is often very well approximated by simpler Markovian dynamics. How this memorylessness emerges is not dissimilar to questions, regarding the emergence of thermodynamic behaviour, which have pervaded quantum mechanics since its conception. Indeed, it can be shown that canonical quantum states are typical~\cite{Normal_typicality, Lloyd_1988, Goldstein_2006, Gemmer_2009}, and we now know that the fundamental postulate of equal a-priori probabilities of statistical mechanics can be traced back to the entanglement between subsystems and their environment~\cite{Popescu2006}. It turns out that, very similarly, if we sample a generic quantum process occurring in a large finite environment at random, it will be almost Markovian with very high probability~\cite{FigueroaRomero2019almostmarkovian}.

This sampling procedure can be formalized through the so-called Haar probability measure, $\mu_\mathsf{h}$, over the $d$-dimensional unitary group $\mbb{U}(d)$, which is the unique (up to a multiplicative constant) measure with the property that, if $U\in\mbb{U}(d)$ is distributed according to the Haar measure, then so is any composition $UV$ or $VU$, with a fixed $V\in\mbb{U}(d)$. It can be normalized to one, so as to constitute a legitimate probability measure~\cite{GuMoments}. The Haar measure allows one to swiftly obtain statistical properties of uniformly distributed quantities~\cite{GuMoments, Collins2006, Puchala_2017, PhysRevLett.71.1291, Giraud_2007, Pasquale_2011} and, furthermore, to prove concentration of measure results~\cite{Ledoux, milman, boucheron2013}; these somewhat surprisingly imply that, when drawn from the right distribution, certain quantities will become overwhelmingly likely to be close to another fixed quantity as the Hilbert space dimension is increased. Henceforth, we write $U\sim\mu_\mathsf{h}$ to refer to $U$ as distributed according to the Haar measure and, similarly, we use $\mbb{P}_\mathsf{h}$ and $\mbb{E}_\mathsf{h}$ to denote probabilities and expectations with respect to the Haar measure.

The Result by Modi et al. on Markovian Typicality~\cite{FigueroaRomero2019almostmarkovian}, which is reproduced in detail in the Methods section on, gives a mathematically sound result of concentration of measure around Markovian processes. However, it assumes a Haar-distributed uniform sampling of unitary dynamics, and we know that nature seldom behaves randomly~\cite{Zanardi_2010, Zanardi_2010_2}. The dynamics of a vast number of physically relevant models can be approximated as Markovian~\cite{schlosshauer2007decoherence}, so can we say that these also satisfy a concentration of measure with respect to Markovianity? 

\begin{figure}[t]
\raggedright
\begin{minipage}{0.5\textwidth}
\begin{tikzpicture}
\node[anchor=south west, inner sep=0] (image) at (0,0) {\includegraphics[width=0.6\textwidth, height=0.525\textwidth]{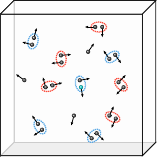}};
\end{tikzpicture}
\end{minipage}%
\caption{\textbf{A toy model analogous to a system with dynamics given by an approximate unitary design with two kinds of two-qubit interactions only.} An impurity particle (teal) immersed in a gas of $n_\mathsf{E}$ particles (arrows depicting direction of motion) within a closed box, where all particles interact in pairs in one of two ways (dashed circles) at random, can be similarly described by an approximate unitary design. The result of Theorem~\ref{Result: Large Dev Markov} ensures that for a large enough $n_\mathsf{E}$ and number of interactions, most processes analogous to this one with approximate unitary designs will be almost Markovian.}\label{fig:gas}
\end{figure}

In some circumstances, sets of physical processes can approximate some of the statistical features of the Haar measure~\cite{Guhr1998, DAlessio2016, Gogolin_2016, mehta2004random}; for example, consider the toy model depicted in Fig.~\ref{fig:gas}, comprising a dilute gas of $n$ particles evolving autonomously in a closed box. The gas particles interact with each other in one of two ways as they randomly move inside the box. Following and intervening on a special impurity particle, taken to be the system, this model can be approximately thought to be described by a circuit such as the one in Fig.~\ref{Fig:Processes}(b). The simplicity of this system suggests that it can only uniformly randomize after a large number of random two-qubit interactions, progressively resembling genuine Haar random dynamics.

One possible way to quantify this progressive resemblance of the Haar measure is given by the concept of unitary designs. In general an $\epsilon$-approximate $\design$-design, which we denote $\mu_{\,\design_\epsilon}$, can be defined through
\begin{gather}
    \left\|\mbb{E}_{\,\design_\epsilon}\left[\mc{V}^{\otimes{s}}(X)\right]-\mbb{E}_\mathsf{h}\left[\mc{U}^{\otimes{s}}(X)\right]\right\|\leq\epsilon,\quad\forall{s}\leq{\design}
    \label{eq: def approx t design generic}
\end{gather}
for a suitable metric $\|\cdot\|$, where $\mc{U}(\cdot):=U(\cdot)U^\dg$ and $\mc{V}(\cdot):=V(\cdot)V^\dg$ are unitary maps with $U,V\in\mbb{U}(d)$. Here, as above, the notation $\mathbb{E}_\Omega$ indicates the expectation value with respect to a given probability measure $\mu_\Omega$, i.e. $V\sim\mu_{\,\design_\epsilon}$ and $U\sim\mu_\mathsf{h}$. That is, $\mu_{\,\design_\epsilon}$ approximates the Haar measure up to the $\design$\textsuperscript{th} moment with a small error $\epsilon$. In the case we are interested in, the unitary maps will correspond to $\mathsf{SE}$ unitaries, as depicted in Fig.~\ref{Fig:Processes}(a), according to the either the Haar measure or a unitary design. We also do not assume anything about the parameter $\design$ other than it is a positive non-zero integer.

Notice what this would mean for a model similar to that of Fig.~\ref{fig:gas}: as individual random two-body interactions of each kind accumulate, what we expect is for the dynamics to start scrambling their information across the whole gas in the box, progressively becoming more complex and uniformly random~\cite{Roberts2017}. Unitary designs give us this finite quantification of the approximation to uniform Haar randomness and, in this case, it can give us a precise way to account for the progressive emergence of complexity from seemingly simple individual two-body interactions.

Unitary designs for $\design=2,3$ have been widely studied~\cite{Emerson_2005, PhysRevA.80.012304, Gross2007, Harrow2009, Dankert_2009, Nakata2013, Wallman_2014, Webb2015, Winter_2017, PhysRevA.96.062336} and efficient constructions are known for larger values of $\design$~\cite{Brandao2016, Winter_HamDesign, Harrow2009}. The latter are of particular relevance, precisely as designs for large $\design$, i.e., those with a higher complexity~\cite{Roberts2017}, are expected to satisfy tighter large deviation bounds, approaching concentration of measure as the level and quality of the design increases.

Such large deviation bounds over approximate unitary designs were derived in a general form by R. Low~\cite{Low_2009} for a polynomial function satisfying a concentration of measure bound, and we now use them to demonstrate the phenomenon of Markovianization for corresponding classes of processes.
\begin{theorem}\label{Result: Large Dev Markov}
Given a $k$-step process $\Upsilon$ on a $d_\mathsf{S}$ dimensional subsystem, generated from global unitary $d_\mathsf{SE}$ dimensional $\mathsf{SE}$ dynamics distributed according to an $\epsilon$-approximate unitary $\design$-design $\mu_{\,\design_\epsilon}$, the likelihood that its non-Markovianity exceeds any $\delta>0$ is bounded as
\begin{gather}
  \mbb{P}_{\design_\epsilon}[\,\mc{N}_\bdiamond\geq\delta]\leq\mathsf{B},
  \label{eq: largedev_design}
\end{gather}
where $\mathsf{B}$ is defined as
\begin{gather}
    \mathsf{B}:=\f{d_\mathsf{S}^{3m(2k+1)}}{\delta^{2m}}\left[\left(\f{m}{\mc{C}}\right)^{m}\hspace{-0.05in}+(2\mc{B})^{2m}+\f{\epsilon}{d_\mathsf{SE}^{\,\design}}\eta^{2m}\right],
\end{gather}
for any $m\in(0,\design/4]$ and
\begin{gather}
    \eta:=\left(d_{\mathsf{SE}}^4d_\mathsf{S}^{2k}+d_\mathsf{S}^{-(2k+1)}\right)/4,
    \label{eq: def eta}
\end{gather}
where $\mc{C}$ is defined in Eq.~\eqref{eq: Lipschitz Haar} and $\mc{B}$ an upper bound on the expected norm-1 non-Markovianity $\mbb{E}_\mathsf{h}[\mc{N}_1]$, defined in Eq.~\eqref{Eq: B Haar upper bound}.
\end{theorem}

The proof is displayed in full in the Methods section. The overall strategy is as done by R. Low~\cite{Low_2009}: a bound on the moments $\mbb{E}_{\,\design_\epsilon}[\mc{N}_\bdiamond^{\,m}]$ is given in terms of $\mc{B}$, $\mc{C}$ and $\eta$, followed by Markov's inequality. The quantity $\eta$ is related to the $\epsilon$-approximate unitary $\design$-design $\mu_{\,\design_\epsilon}$ through
\begin{align}
    \mbb{E}_{\,\design_\epsilon}\left[\mc{N}_2^{\,2m}\right]\leq\mbb{E}_\mathsf{h}\left[\mc{N}_2^{\,2m}\right]+\f{\epsilon}{d_\mathsf{SE}^{\,\design}}\,\eta^{2m},
    \label{eq: Low design eta} 
\end{align}
for any $m>0$ and corresponds to the sum of the moduli of the coefficients of $\mc{N}_2^{\,2}$. We explicitly determine a bound on this quantity within the proof of Theorem~\ref{Result: Large Dev Markov} in the Methods section, which is the one we take as definition in Eq.~\eqref{eq: def eta}.

\begin{figure}[t]
\raggedleft
\includegraphics[width=0.495\textwidth]{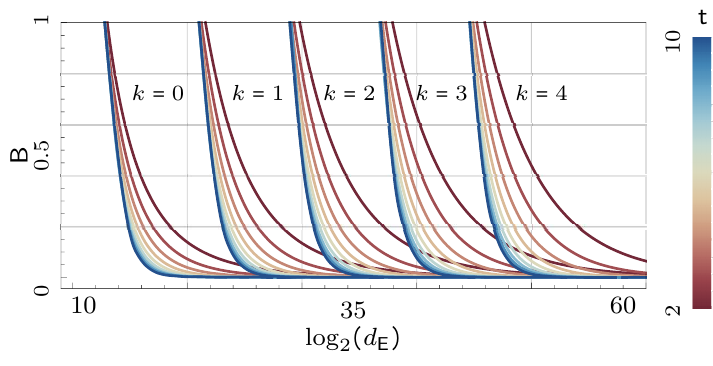}
\caption{\textbf{Upper bound on the probability for non-Markovianity to exceed a small amount for processes with distinct number of interventions and design dynamics against environment size }. Upper bound $\mathsf{B}$, defined by Eq.~\eqref{eq: largedev_design}, on $\mbb{P}_{\design_\epsilon}[\mc{N}_\bdiamond\geq0.1]$, the probability $\mbb{P}_{\design_\epsilon}$ over an $\epsilon$-approximate $\design$-design for the non-Markovianity $\mc{N}_\bdiamond$ to exceed $\delta=0.1$, against $\log_2(d_\mathsf{E})$, where $d_\mathsf{E}$ is environment dimension, for a subsystem qubit undergoing a joint closed approximate unitary design interaction between a given number of interventions $k$. We fix an $\epsilon=10^{-12}$ approximate unitary $\design$-design for different values $2\leq{\design}\leq10$ and fixed values of timesteps $k$, optimizing $m$ for each case.}
  \label{Fig: plot Prob-dE}
\end{figure}

The choice of $0<m\leq{\design}/4$ can be made to optimize the right-hand-side of the inequality, which ideally should be small whenever $\delta$ is. The term $d_\mathsf{S}^{3(2k+1)}/\delta^2$ arises from bounding $\mc{N}_\bdiamond$ and Markov's inequality, while the three summands within square brackets will be small provided \emph{i}) $\mc{C}$ is large, \emph{ii}) $\mc{B}$ is small and \emph{iii}) the unitary design sufficiently small $\epsilon$ and large $\design$ {is well-approximate and high enough}. For conditions \emph{i}) and \emph{ii}), we require a fixed $k$ such that $d_\mathsf{E}\gg{d}_\mathsf{S}^{2k+1}$: this implies $\mc{B}\approx0$, so that ignoring subleading terms, we require $\epsilon\ll\delta^{2m} \left(2d_\mathsf{E}^{-2}d_\mathsf{S}^{-(10k+11)/4}\right)^{4m}d_\mathsf{SE}^{\,\design}$ for a meaningful bound, as detailed in the Methods section on Convergence towards Markovianity.

Overall, the bound in Eq.~\eqref{eq: largedev_design} approaches concentration whenever $d_\mathsf{E}$ is large relative to $d_\mathsf{S}$ and $k$, together with large enough $\design$, as shown in Fig.~\ref{Fig: plot Prob-dE}. Generally, it can be seen by inspection that the scaling in these cases will be polynomially vanishing in $d_\mathsf{E}$, exponentially vanishing in $\design$ (upon appropriate choice of parameter $m$), and becomes loose, polynomially in $d_\mathsf{S}$ and exponentially in $k$. Therefore, the vast majority of processes sampled from such a $\design$-design are indistinguishable from Markovian ones in this limit. This can be intuitively understood as that for processes of small subsystems in large environments ($d_\mathsf{E}\gg{d}_\mathsf{S}^{2k+1}$) undergoing complex enough dynamics (large enough $\design$) will look almost Markovian with high probability if the system is probed not too many times (small $k$). We will now show how these processes can be modelled in terms of random circuits.

\subsection{Markovianization by circuit design}
While no explicit sets forming unitary $\design$-designs for $\design\ge4$ are known to date, several efficient constructions generating approximate unitary designs by quantum circuits are known. Using these constructions we can highlight the physical implications of the theorem above. We begin by discussing the details of one such construction. As suggested in Fig.~\ref{Fig:Processes}(b), this construction only requires simple two-qubit interactions and, under certain conditions, yields an approximate unitary design, from which we can use Eq.~\eqref{eq: largedev_design} in our main Theorem to verify that Markovianization emerges.

\begin{figure}[t]
\centering
\includegraphics[width=0.48\textwidth]{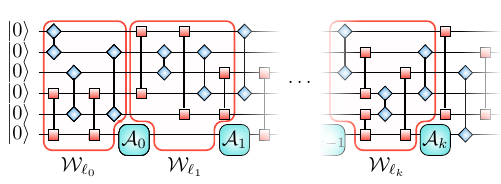}
  \caption{\textbf{Circuit diagram for a quantum process which can Markovianize under only two different types of 2-qubit interaction dynamics.} For an $n$-qubit system (where each $|0\rangle$ is a single qubit), the unitaries $\mc{W}_\ell$, composed of $\ell$ alternate repetitions only two distinct types of random interactions (depicted by diamonds and squares joined by the interacting qubits), and defined by Eq.~\eqref{eq: circuit W}, generate an $\epsilon$-approximate unitary $\design$-design whenever $\ell\geq{\design}-\log_2(\epsilon)/n$, as shown bt Winter et al.~\cite{Winter_HamDesign}. This can be thought as stemming from repeated alternate applications of random 2-qubit gates diagonal in only two Pauli bases. A qubit probed with a set of operations $\{\mc{A}_i\}$ on a system undergoing $\epsilon$-approximate unitary $\design$-design dynamics $\mc{W}_\ell$ on a large environment will Markovianize for small design error $\epsilon$ and large complexity $\design$ as specified in the main text.}
  \label{fig: RDC}
  \end{figure}

We focus specifically on Result 2 by Winter et al.~\cite{Winter_HamDesign}, reproduced in the Methods section on efficient unitary designs, where a circuit with interactions mediated by two-qubit diagonal gates with three random parameters is introduced. The intuition behind such construction is that repeated alternate applications of these diagonal gates quickly randomizes the system. Notice that this idea now fully captures the gas scenario depicted in Fig.~\ref{fig:gas}, where we only have two types of random two-body interactions repeatedly occurring, and we focus on one of the particles of the gas. The detail of this construction is reproduced in the Methods section on efficient circuit unitary designs.

We can illustrate this idea in Fig.~\ref{fig: RDC}, where we depict an $n$-qubit $\mathsf{SE}$ composite with $k$ interventions on one of the qubits, with the interactions within the circuit being only between pairs of qubits and of only two kinds; these form blocks of unitaries between each time-step $i$ that we label $\mc{W}_{\ell_i}$, where $\ell$ is related to the amount of two-qubit interactions as explicitly defined in Eq.~\eqref{eq: circuit W}. The main Result 2 of by Winter et al.~\cite{Winter_HamDesign} states that for an $n$-qubit system, when $\design$ is of order $\sqrt{n}$, a circuit $\mc{W}_\ell$ yields an $\epsilon$-approximate unitary $\design$-design if $\ell\geq{\design} - \log_2(\epsilon)/n$, up to leading order in $n$ and $\design$.

Furthermore, of great relevance in this result is the fact that almost all 2-qubit gates in each repetition of $\mc{W}_\ell$ can be applied simultaneously because they commute~\cite{Nakata2017decouplingrandom, Winter_2017}. Therefore, if $\mc{W}_\ell$ yields an approximate unitary design as above, the order of the non-commuting gate depth $\mathfrak{D}$, defined by Winter et al.~\cite{Nakata2017decouplingrandom} as the circuit depth when each commuting part of the circuit is counted as a single part, will coincide with the bound on the order of the number of repetitions $\ell$. That is, the non-commuting gate depth asymptotes to
\begin{gather}\label{eq:depth}
    \mathfrak{D}\sim{\design}-\log_2(\epsilon)/n.
\end{gather}

\begin{figure}[t]
\raggedright
\includegraphics[width=0.495\textwidth]{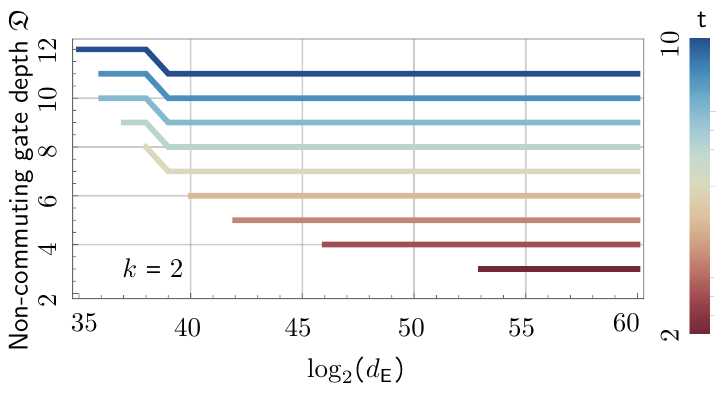}
\caption{\textbf{Scaling of the non-commuting gate depth of the approximate unitary design by Winter et al.~\cite{Winter_HamDesign} for a 2-step process on a single qubit to Markovianize with respect to environment size.} Scaling of the non-commuting gate depth $\mathfrak{D}$, given by the minimum amount of alternate repetitions $\ell$ of the two kinds of random two-qubit diagonal gates within the unitary $\mc{W}_\ell$, plotted against the environment qubits  $n_\mathsf{E}=\log_2(d_\mathsf{E})$, to generate an $\epsilon=10^{-12}$ approximate unitary $\design$-design for $2\leq{\design}\leq10$. This is such that for a single-qubit system undergoing a process with $k=2$ timesteps, the probability $\mbb{P}_{\design_\epsilon}$ for the non-Markovianity $\mc{N}_\bdiamond$ exceeding $0.1$ is less or equal than $0.01$, i.e. $\mbb{P}_{\design_\epsilon}[\mc{N}_\bdiamond\geq0.1]\leq\mathsf{B}\leq0.01$.}\label{Plot: nqubits}
\end{figure}

We can now think of the system from the toy model of Fig.~\ref{fig:gas} as given by a spin locally interacting with a large, $n_\mathsf{E}$-qubit environment, via a random time-independent Hamiltonian, with Eq.~\eqref{eq: largedev_design} statistically predicting under which conditions memory effects can be neglected. Notice that this is only a physical picture evoked by the $\mc{W}_\ell$ circuits rather than exactly being the model described by it. In Fig.~\ref{Plot: nqubits} we take such a system for a single qubit and demand a bound $\mathsf{B}\leq0.01$ on the probability $\mbb{P}_{\design_\epsilon}[\mc{N}_\bdiamond\geq0.1]$ for a $k=2$ timestep process; with this, we plot the scaling of the non-commuting gate depth $\mathfrak{D}$ required to achieve an $\epsilon=10^{-12}$ approximate unitary $\design$-design using $\mc{W}_\ell$ circuits for different values of $2\leq{\design}\leq10$. While the number of 2-qubit gates is on the order of $10^4$, the number of repetitions $\ell$ is at most 12 for an approximate 10-design and stays mostly constant as the number of environment qubits increases.

This construction naturally accommodates the cartoon example in Fig.~\ref{fig:gas}. As long as the two interactions in the example together generate the necessary level of complexity, Markovianization will emerge. This shows, in principle, how simple dynamics described by approximate unitary designs can Markovianize under the right conditions. Moreover, taking the physical interpretation of a qubit locally interacting through two-qubit diagonal unitaries with a large environment, it also hints at how macroscopic systems can display Markovianization of small subsystem dynamics in circuits requiring just a small gate depth. Furthermore, for macroscopic systems with coarse observables, the same Markovianization behaviour would remain resilient to a much larger number of interventions.

\section{Discussion}
We have shown how physical quantum processes Markovianize, i.e. forget the past, for a class of physically motivated systems that can finitely can approximate random ones. Forgetfulness is indeed a common feature of the world around us, and one that is crucial for doing science. Without forgetfulness, repeatability would be impossible. After all, if each carbon atom remembered its own past then it will be unique and there would be no sense in classifying atoms and molecules. Beyond these foundational considerations, our results have direct consequences for the study of open systems using standard tools, such as master equations and dynamical maps. The latter of which can be seen as a family of one-step process tensors (with initial $\mathsf{SE}$ correlations a minimum of two steps must be considered~\cite{PhysRevLett.114.090402, Vega_2020}). Specifically, our results, for the case of $k\le 2$, can be used to estimate the time scale, using gate depth as a proxy, on which an approximate unitary design's open dynamics can be described (with high probability) with a truncated memory kernel~\cite{breuer2002theory, Cerrillo2014, Pollock2018tomographically}, or even a Markovian master equation.

Conversely, for larger $k$, our results would have implications for approximations made in computing higher order correlation functions, such as the quantum regression theorem~\cite{Guarnieri2014}. These higher order approximations are independent of those at the level of dynamical maps, which can, e.g., be divisible, even when the process is non-Markovian~\cite{Milz_CPdiv}. This is reflected in the loosening behaviour of the bound in Eq.~\eqref{eq: largedev_design} as the number of timesteps increases, which can be interpreted as a growing potential for temporal correlations to become relevant when more information about the process is accessible.

This breadth of applicability is in contrast with the results of Modi et al.~\cite{FigueroaRomero2019almostmarkovian}, where it was shown that quantum processes satisfy a concentration of measure with respect to Haar measure around Markovian ones, which has two main drawbacks: first, as stated above, Haar random interactions do not exist in nature and hence the relevance of the result is limited. Second, the rate of Markovianization is far too strong. Almost all processes, sampled according to the Haar measure, will simply look random, i.e., Markov order $m=0$ even for a large $k$. This, unlike our current result, misses almost all interesting physical dynamical processes. While the behaviour of our large deviation bound is polynomial, rather than exponential, thus not exhibiting concentration per-se, we have nevertheless exemplified how, with modestly large environments and relatively simple interactions, almost Markovian processes can come about with high probability. Physical macroscopic environments will be far larger than the scale shown in Figs.~\ref{Fig: plot Prob-dE}~and~\ref{Plot: nqubits}.

Despite the fundamental relevance of our result, it is well known that typicality arguments can have limited reach. For instance, the exotic Hamiltonians, introduced by Gemmer et al.~\cite{Gemmer_2020}, which lead to strange relaxation, may not Markovianize even though the $\mathsf{SE}$ process is highly complex with a large $\mathsf{E}$. There is also still significant scope for further addressing physical aspects, such as the question of whether, and how, a time-independent Hamiltonian can give rise to an approximate unitary design~\cite{Winter_HamDesign}, the relevant time scales of Markovianization, or the potential role of different approaches to pseudo-randomness such as that by Kastoryano et al.~\cite{PhysRevX.7.041015}, where it is shown that driven quantum systems can converge rapidly to the uniform distribution. Furthermore, a renewed wave of interest in thermalization has come along with the so-called Eigenstate Thermalization Hypothesis (ETH), which is a stronger and seemingly more fundamental condition on thermalization~\cite{Deutsch_ETH,Srednicki_1994, Srednicki_1999,DAlessio_2016,Murthy_2019,Brenes_2020}, and we would thus expect a deep connection in the sense of ETH between Markovianization and thermalization to be forthcoming. In any case, it is clear that many physical systems Markovianize at some scale, and it only remains to discover how.

\section{Methods}\label{Sec: Methods}
\subsection{The process tensor}\label{Methods: The process tensor}
The Choi state representation of the process tensor is given by
\begin{gather}
    \Upsilon=\tr_\mathsf{E}\left[\,\mathsf{U}_{k:0}\left(\rho\otimes\Psi^{\otimes{k}}\right)\,\mathsf{U}^\dg_{k:0}\right],
    \label{eq: Upsilon}
\end{gather}
where each $\Psi$ is a maximally entangled state on an ancillary space of dimension $d_\mathsf{S}^2$, and where
\begin{gather}\mathsf{U}_{k:0}:=(U_k\otimes\mbb1)\mc{S}_k\cdots(U_1\otimes\mbb1)\mc{S}_1(U_0\otimes\mbb1),\end{gather}
is a unitary operator acting on the whole $\mathsf{SE}$ together with the $2k$ ancillas. All identities act on the ancillary system, the $U_i$ are $\mathsf{SE}$ unitary operators at step $i$, and $\mc{S}_i$ is a $\swap$ operator between system $\mathsf{S}$ and half of the $i$\textsuperscript{th} ancillary space at the $i$th time-step of the process.

The definition in Eq.~\eqref{eq: Upsilon} is a generalization of the standard Choi state for quantum channels, as given by the Choi-Jamio\l{}kowski isomorphism (CJI)~\cite{watrous_2018}. The CJI for quantum channels establishes a one to one correspondence with a quantum state on a larger Hilbert space, given as the action of the channel onto half a maximally entangled state. The standard definition uses unnormalized maximally entangled states; however, here we are concerned with distinguishability of Choi states through the diamond norm in Eq.~\eqref{Eq: def nM diamond} and the Schatten norms in Eq.~\eqref{eq: Schatten}, so we avoid a normalization factor in these by normalizing the Choi states by definition. A discussion in full depth about the process tensor, its different representations and its properties and relevance is given by Modi et al.~\cite{Simon_Operational}.

As stated in the main text, $\Upsilon$ can be efficiently described when written as a matrix product operator~\cite{PhysRevA.97.012127, Verstraete_2004}. A matrix product operator (\textsf{MPO}) gets its name from the representation of an $n$-body operator $\hat{O}$ as ${\hat{O}=\sum_{\{p,q\}} O_{p_1\ldots p_n}^{q_1 \ldots q_n}|p_1\ldots p_n\rangle\!\langle q_1\ldots q_n|}$, where the coefficients can be represented as a product of matrices, ${O_{p_1\ldots p_n}^{q_1 \ldots q_n}=\tr[M_1^{p_1q_1}M_2^{p_2q_2}\cdots M_n^{p_n q_n}]}$. In particular, a matrix product density operator is a \textsf{MPO} with $M_i^{p_iq_i}=\sum_\ell A_i^{p_i\ell}\otimes(A_i^{q_i\ell})^\dg$, with the dimension of the matrices $A_i^{p_i\ell}$ known as the bond dimension. For the process tensor $\Upsilon$, these matrices generically correspond to ${M_i^{r_ir^\prime_{i-1}s_i s^\prime_{i-1}}=\langle{r}_i|U_i|r^\prime_{i-1}\rangle\otimes\langle{s}_i|U_i^\dg|s^\prime_{i-1}\rangle}$ where ${|r^{(\prime)}\rangle}$ and ${|s^{(\prime)}\rangle}$ are subsystem $\mathsf{S}$ basis vectors and $U_i$ is an $\mathsf{SE}$ unitary at timestep $i$~\cite{PhysRevA.97.012127}. This means the bond dimension of $\Upsilon$ is $d_\mathsf{E}$, which in practice should be much smaller, given that only part of
the environment interacts with the system at any given time.

\subsection{A non-ambiguous measure of non-Markovianity}
As with any distinguishability measure, the non-Markovianity metric of Eq.~\eqref{Eq: def nM diamond} is not unique, and we choose the diamond norm for its mentioned operational significance. However, more generally, for any Schatten $p$-norm $\|X\|_p:=\tr(|X|^p)^{\frac{1}{p}}$, a similar quantity can be defined $\mc{N}_p:= \f{1}{2} \min_{\Upsilon^\markov} \|\Upsilon-\Upsilon^\markov\|_p,\label{eq: Schatten}$ as done with $p=1$ in by Modi et al.~\cite{FigueroaRomero2019almostmarkovian}, whenever $\Upsilon$ is normalized such that $\tr[\Upsilon]= \tr[\Upsilon^\markov]=1$. Then, we have the hierarchy $\mc{N}_1\geq\mc{N}_2\geq\ldots$, induced by that of the Schatten norms. As the black diamond norm is generally difficult to compute exactly, a particularly useful relation to Eq.~\eqref{Eq: def nM diamond} is $d_\mathsf{S}^{-2k-1} \mc{N}_\bdiamond \leq \mc{N}_1\leq\mc{N}_\bdiamond$, in the sense that once any Schatten norm is known, the black diamond norm is automatically bounded.

Nevertheless, we highlight that, in general, any distinguishability measure $\mc{N}$ between a process $\Upsilon$ and the closest Markovian one $\Upsilon^\markov$ will capture all non-Markovian features across multiple time steps, i.e., all multi-time phenomena and memory effects~\cite{PhysRevLett.120.040405}. This is in contrast to other measures of non-Markovianity, e.g. trace-distance based measure~\cite{RevModPhys.88.021002} and other based on divisibility~\cite{Rivas_2014}, that have been proposed in recent years. In particular, all other measures relying on completely positive divisibility are only able to account for temporal correlations across at most three time-steps and are not sufficient to enforce the multi-time Markov condition~\cite{PhysRevLett.123.040401}. This is even true in the classical case. Concretely, there are explicit examples of multi-time non-Markovian processes that are shown to be completely positive divisible processes, thus also deemed to be Markovian by the trace-distance based measure~\cite{PhysRevLett.120.040405, PhysRevLett.123.040401}. On the other hand, if a process satisfies the multi-time Markov condition, then it will be completely positive divisible.

In other words, the multi-time Markov condition is a stronger one that contains Markov conditions based on completely positive divisibility. This is why we consider the multi-time Markov condition in this manuscript.

\subsection{Markovian Typicality}\label{Methods: Markovian typicality}
In general, we say that a function $f$ from a metric space $\mathfrak{S}$ with metric $\Delta_\mathfrak{S}$ and probability measure $\mu_\sigma$, to the real numbers, satisfies a concentration of measure around its mean if, for any point $x\in\mathfrak{S}$ and any $\delta>0$,
\begin{gather}
    \mbb{P}_\sigma[f(x)\geq\mbb{E}_\sigma(f)+\delta]\leq\alpha_\sigma(\delta/\mathrm{L}),
    \label{eq concentration of measure}
\end{gather}
where as done in the remainder of this manuscript, $\mbb{P}_\sigma$ and $\mbb{E}_\sigma$ explicitly refer to the probability and expectation with $x\sim\mu_\sigma$, and where $\mathrm{L}>0$ is the so-called Lipschitz constant of $f$, which can be determined according to $|f(x)-f(y)| \leq \mathrm{L} \, \Delta_\mathfrak{S}(x,y)$ for any two points $x,y\in\mathfrak{S}$. Whenever $\mathrm{L}$ is small, intuitively this implies that $f$ varies slowly in such space. Finally, the function $\alpha_\sigma$ is called a concentration rate; it generally must be vanishing in increasing $\delta$ in order for \ref{eq concentration of measure} to constitute concentration of measure, and it intuitively tells us how strong such concentration is.

Particularly well-known is the example of concentration of measure in the hypersphere of a high dimension, where for all functions that do not change too rapidly, i.e. with a small Lipschitz constant $\mathrm{L}$, the function evaluated on a point picked uniformly at random will be close to its mean value with high probability, i.e. specifically $\alpha_\sigma$ decays exponentially with $-\delta^2$. This is also known as Levy's lemma~\cite{Ledoux} and it has, remarkably, also been used by Winter et al.~\cite{Popescu2006} to show that the fundamental theorem of statistical mechanics arises from entanglement.

Similarly, Modi et al.~\cite{FigueroaRomero2019almostmarkovian} showed that quantum processes satisfy a concentration of measure around Markovian ones, explaining the emergence of Markovianity without a-priori assumptions. In particular, there, the trace distance $\mc{N}_1$ was used as a measure of non-Markovianity, which strictly speaking gives the distinguishability between explicitly constructed Choi states of corresponding process tensors and has no operational meaning; however, we can use the relation $d_\mathsf{S}^{-2k-1} \mc{N}_\bdiamond\leq\mc{N}_1\leq\mc{N}_\bdiamond$ to relate this to the stricter notion of non-Markovianity defined in terms of the diamond norm in Eq.~\eqref{Eq: def nM diamond}. This implies that the main result by Modi et al.~\cite{FigueroaRomero2019almostmarkovian}, where all $\mathsf{SE}$ unitaries of Eq.~\eqref{eq: Upsilon} were randomly sampled according to the Haar measure, can be written equivalently as
\begin{gather}
    \mathbb{P}_\mathsf{h}\left[\mc{N}_\bdiamond\geq{d}_\mathsf{S}^{2k+1}\mc{B}+\delta\right]\leq \exp\left\{ -4\,\mc{C}\,\delta^2 d_\mathsf{S}^{-2(2k+1)} \right\},
    \label{eq: non-Markov concentration Haar}
\end{gather}
where
\begin{gather}
    \mc{C}=\f{d_\mathsf{SE}(k+1)}{16} \left(\f{d_\mathsf{S}-1}{d_\mathsf{S}^{k+1}-1}\right)^2,
    \label{eq: Lipschitz Haar}
\end{gather}
is the Lipschitz constant of $\mc{N}_1$, and
\begin{gather}
    \mc{B}=\begin{cases}
    \displaystyle{\f{1}{2}\sqrt{d_\mathsf{E}\,\mathbb{E}_\mathsf{h}[\tr(\Upsilon^2)]-x}+\f{y}{2}}
    &\text{if}\quad d_\mathsf{E}<d_\mathsf{S}^{2k+1}
    \\[0.2in]
    \displaystyle{\f{1}{2}\sqrt{d_\mathsf{S}^{2k+1}\mathbb{E}_\mathsf{h}[\tr(\Upsilon^2)]-1}}&\text{otherwise},\end{cases}
    \label{Eq: B Haar upper bound}
\end{gather}
is an upper bound on $\mbb{E}_\mathsf{h}[\mc{N}_1]$, the expected non-Markovianity over the Haar measure, with $x:= d_\mathsf{E}d_\mathsf{S}^{-(2k+1)}\left(1+y\right)$ and $y := 1-d_\mathsf{E}d_\mathsf{S}^{-(2k+1)}$, and
\begin{gather}
    \mbb{E}_\mathsf{h}[\tr(\Upsilon^{2})]=\f{d_\mathsf{E}^2-1}{d_\mathsf{E}(d_\mathsf{SE}+1)}\left(\f{d_\mathsf{E}^2-1}{d_\mathsf{SE}^2-1}\right)^k+\f{1}{d_\mathsf{E}},
\end{gather}
the expected purity, i.e. the noisiness of the process $\Upsilon$, over the Haar measure. Holding everything else constant, the bound $\mc{B}\geq\mbb{E}_\mathsf{h}[\mc{N}_1]$ satisfies
\begin{gather}
    \lim_{d_\mathsf{E}\to\infty}\mc{B}=0\quad \text{and} \quad \lim_{k\to\infty}\mc{B}=1,
\end{gather}
so that the expected non-Markovianity vanishes in the $d_\mathsf{E}\to\infty$ limit and becomes loosest in the $k\to\infty$ limit case.

The significance of Eq.~\eqref{Eq: B Haar upper bound} is thus that quantum processes with not too many interventions in high dimensional environments will look to be almost Markovian with high probability. This means that, even when processes generically carry temporal correlations, these are typically low, explaining the emergence of Markovian processes without ad-hoc assumptions such as the Born-Markov approximation of weak coupling~\cite{schlosshauer2007decoherence}.

\subsection{Unitary designs}
The result in Eq.~\eqref{Eq: B Haar upper bound} assumes that the dynamics are Haar distributed; however, implementing a Haar random unitary requires an exponential number of two-qubit gates and random bits~\cite{Knill_1995}, thus Haar random dynamics cannot be obtained efficiently in a physical setting.

An exact unitary $\design$-design is defined~\cite{Low_2009} as a probability measure $\mu_{\,\design}$ on $\mbb{U}(d)$ such that for all positive $s\leq{\design}$, and all $d^s\times{d}^s$ complex matrices $X$,
\begin{gather}
    \mbb{E}_{\,\design}\left[\mc{V}^{\otimes{s}}(X)\right]=\mbb{E}_\mathsf{h}\left[\mc{U}^{\otimes{s}}(X)\right],\quad\forall{s\leq{\design}}.
    \label{eq: def exact t design}
\end{gather}

As per the definition in Eq.~\eqref{eq: def exact t design}, a unitary $\design$-design reproduces up to the $\design$\textsuperscript{th} moment over the uniform distribution given by the Haar measure. In particular, $\mu_{\,\design}$ can consist of a finite ensemble $\{V_i,p_i\}_{i=1}^N$ of unitaries $V_i$ and probabilities $p_i$, as is now common in applications such as so-called randomized benchmarking of error rates in quantum gates~\cite{Dankert_2009, Wallman_2014}.

Moreover, this definition can be relaxed by letting a unitary design approximate the Haar measure with a small error $\epsilon$. In this manuscript we specifically employ the definition by R. Low~\cite{Low_2009} for unitary designs. It uses the fact that the definition of an exact $\design$-design, $\mu_{\,\design}$, can be written in terms of a balanced monomial $\Theta$ of degree less or equal to $\design$ in the elements of the unitaries $U$. A balanced monomial of degree $\design$ is a monomial in the unitary elements with precisely $\design$ conjugated and $\design$ unconjugated elements: for example, $U_{ab}U_{cd}U_{ef}^*U_{hg}^*$ is a balanced monomial of degree 2. Thus, writing Eq.~\eqref{eq: def exact t design} in terms of matrix elements, this can be seen to be equivalent to requiring $\mbb{E}_{\,\design}[\Theta(V)]=\mbb{E}_\mathsf{h}[\Theta(U)]$ for all monomials $\Theta$ of degree $s\leq{\design}$. Similarly, for an $\epsilon$-approximate $\design$-design we adopt the definition by R. Low~\cite{Low_2009} with Eq.~\eqref{eq: def approx t design generic} implying
\begin{gather}
	\left|\mbb{E}_{\,\design_\epsilon}\Theta(V)-\mbb{E}_\mathsf{h}\Theta(U)\right|\leq\f{\epsilon}{d^{\,\design}},
	\label{eq: approximate design Low}
\end{gather}
for monomials $\Theta$ of degree $s\leq{\design}$. From now on, we will focus on the more general approximate designs. We will see below that the degree $\epsilon$ to which the distribution of the unitary dynamics on $\mu_{\,\design_\epsilon}$ differs from an exact design for given $\design$ depends on the complexity of the model.

\subsection{Large deviation bounds for \texorpdfstring{$\design$}{t}-designs}\label{Appendix: Low Result}
The general idea for the main result by R. Low~\cite{Low_2009} (similarly applied before by Horodecki et al.~\cite{Brandao2016}) is that given a $\mu_{\,\design_\epsilon}$ distribution as an $\epsilon$-approximate unitary $\design$-design and a concentration result for a polynomial $\mc{X}$ of degree $p$, then one can compute the last term $f_{\,\design_\epsilon}$ in
\begin{gather}
	\mbb{E}_{\,\design_\epsilon}\mc{X}^m=\mbb{E}_\mathsf{h}\mc{X}^m+f_{\,\design_\epsilon},
\end{gather}
with $m\leq{\design/2p}$, which will generally have a dependence $f_{\,\design_\epsilon}=f_{\,\design_\epsilon}(\epsilon,\design,\mc{X})$. Using Markov's inequality
\begin{align}
	\mbb{P}_{\design_\epsilon}(\mc{X}\geq\delta)&=\mbb{P}_{\design_\epsilon}(\mc{X}^m\geq\delta^m)\nonumber\\
	&\leq\f{\mbb{E}_{\,\design_\epsilon}\mc{X}^m}{\delta^m}\nonumber\\
	&=\f{1}{\delta^m}\left[\mbb{E}_\mathsf{h}\mc{X}^m+f_{\,\design_\epsilon}\right],
\end{align}
which is the form of the main large-deviation bound.

Specifically, the results that we employ are the following, proved R. Low~\cite{Low_2009}.

\begin{theorem}[Large deviation bounds for $\design$-designs by R. Low\cite{Low_2009}]
\label{thm: large dev Low}
Let $\mc{X}$ be a polynomial of degree $\mathsf{T}$. Let $f(U)=\sum_i\alpha_i\Theta_{s_i}(U)$ where $\Theta_{s_i}(U)$ are monomials and let $\alpha(f)=\sum_i|\alpha_i|$. Suppose that $f$ has probability concentration
\begin{equation}
    \mbb{P}_\haar[|f-\zeta|\geq\delta]\leq C\exp\left(-C\delta^2\right),
\end{equation}
and let $\mu_{\design_\epsilon}$, be an $\epsilon$-approximate unitary $\design$-design, then
\begin{equation}
    \mbb{P}_{\mu_{\design_\epsilon}}[|f-\zeta|\geq\delta]\leq\f{1}{\delta^{2m}}\left(C\left(\f{m}{C}\right)^m+\f{\epsilon}{d^{\,\design}}(\alpha+|\zeta|)^{2m}\right),
    \label{eq: Low main large dev}
\end{equation}
for any integer $m$ with $2m\mathsf{T}\leq\design$.
\end{theorem}

This is the most general result providing a large-deviations bound on approximate unitary designs, where $\zeta$ can be any quantity, in particular the expectation of $f$. The main idea from this result (similarly applied before by Horodecki et al.~\cite{Brandao2016}) is that given a $\mu_{\mathsf{t}_\epsilon}$ distribution as an $\epsilon$-approximate unitary $\design$-design and a concentration result for a polynomial $f$ of degree $\mathsf{T}$, then one can compute
\begin{gather}
	\mbb{E}_{\,{\mathsf{t}_\epsilon}}\left[f^m\right]=\mbb{E}_\haar\left[f^m\right]+g(\epsilon,\design,f),
\end{gather}
where $m\leq{\design/2\mathsf{T}}$. Using Markov's inequality we have
\begin{align}
	\mbb{P}_{\mathsf{t}_\epsilon}(f\geq\delta)=\mbb{P}_{\mathsf{t}_\epsilon}{\delta^m}\left[\mbb{E}_\haar\left[f^m\right]+g(\epsilon,\mathsf{t},f)\right],
	\label{eq: large dev form}
\end{align}
which is the form of the main large deviations bound in Eq.~\eqref{eq: Low main large dev}. More precisely, the other two main results that come along with the proof of Theorem~\ref{thm: large dev Low} by R. Low~\cite{Low_2009}, and allowing to compute the right hand-side of Eq.~\eqref{eq: large dev form} are the following.

\begin{lemma}[3.4 of by R. Low~\cite{Low_2009}]
\label{Lemma: Low 1}
Let $\mc{X}$ be a polynomial of degree $\mathsf{T}$ and $\zeta$ any constant. Let $f(U)=\sum_i\alpha_i\Theta_{s_i}(U)$ where $\Theta_{s_i}(U)$ are monomials and let $\alpha(f)=\sum_i|\alpha_i|$. Then for an integer $m$ such that $2m\mathsf{T}\leq\design$ and $\mu_{\mathsf{t}_\epsilon}$ an $\epsilon$-approximate unitary $\design$-design,
\begin{equation}
    \mbb{E}_{\,\mathsf{t}_\epsilon}\left[|f-\zeta|^{2m}\right]\leq\mbb{E}_\haar\left[|f-\zeta|^{2m}\right]+\f{\epsilon}{d^{\,\design}}\left(\alpha+|\zeta|\right)^{2m}.
    \label{eq: Low lemma 1}
\end{equation}
\end{lemma}

\begin{lemma}[5.2 of by R. Low~\cite{Low_2009}]
\label{Lemma: Low 2}
Let $X$ be any non-negative random variable with probability concentration
\begin{equation}
    \mbb{P}(X\geq\delta+\gamma)\leq{C}\exp(-\mathfrak{C}\,\delta^2),
\end{equation}
where $\gamma\geq0$, then
\begin{equation}
    \mbb{E}[X^m]\leq{C}\left(\f{2m}{\mathfrak{C}}\right)^{m/2}+(2\gamma)^m,
    \label{eq: Low lemma 2}
\end{equation}
for any $m>0$.
\end{lemma}

So, in essence, given these results, we determine the right-hand sides of Eq.~\eqref{eq: Low lemma 1} and Eq.~\eqref{eq: Low lemma 2} through the measure of non-Markovianity in Eq.~\eqref{Eq: def nM diamond} and all the other relevant quantities in such terms.

\subsection{Proof of Theorem~\ref{Result: Large Dev Markov}}
\subsubsection{A bound on the Haar moments of \texorpdfstring{$\mc{N}_2$}{N\_2}}
Let us start by noticing that $\|X\|_1\geq\|X\|_2$, so a concentration for $\mc{N}_1$ given by $\mbb{P}_\mathsf{h}[\mc{N}_1\geq\mc{B}+\delta]\leq\exp(-\mc{C}\delta^2)$ where here $\mc{C}=\f{d_\mathsf{SE}(k+1)}{4} \left(\f{d_\mathsf{S}-1}{d_\mathsf{S}^{k+1}-1}\right)^2$ (here 4 times the one defined in Eq.~\eqref{eq: Lipschitz Haar} in the main text), and $\mc{B}$ is defined in Eq.~\eqref{Eq: B Haar upper bound}, also implies
\begin{gather}
    \mathbb{P}_\mathsf{h}[\mc{N}_2\geq\mc{B}+\delta]\leq\mathrm{e}^{-
\mc{C}\delta^2},
\end{gather}
so that in turn Lemma~\ref{Lemma: Low 2} through Eq.~\eqref{eq: Low lemma 2} implies that
\begin{align}
	\mbb{E}_\mathsf{h}[\mc{N}_2^{2m}]&\leq\left(\f{4m}{\mc{C}}\right)^{m}+(2\mc{B})^{2m}\nonumber\\
	&=\left[\f{16m}{(k+1)d_\mathsf{SE}}\left(\f{d_\mathsf{S}^{k+1}-1}{d_\mathsf{S}-1}\right)^2\right]^{m}+(2\mc{B})^{2m},
\end{align}
for any $m>0$.

\subsubsection{A bound on the design moments of \texorpdfstring{$\mc{N}_2$}{N\_2}}\label{appendix: bound eta}
For the case of all unitaries at each step being independently sampled, $\mc{N}_2^{\,2}$ is a polynomial of degree $p=2$ when the unitaries are all distinct (random interaction type). We can thus take $\mc{N}_2^{\,2}$ and apply Lemma~\ref{Lemma: Low 1} for a unitary $\design$-design $\mu_{\,\design_\epsilon}$ with $t\geq{4m}$, which actually holds for real $m>0$, as
\begin{align}
    \mbb{E}_{\,\design_\epsilon}[\mc{N}_2^{\,2m}]\leq\mbb{E}_\mathsf{h}[\mc{N}_2^{\,2m}]+\f{\epsilon}{d_\mathsf{SE}^{\,\design}}\,\eta^{2m}
\end{align}
where $\eta$ is the sum of the moduli of the coefficients of
\begin{align}
    \mc{N}_2^{\,2}&=\left(\f{1}{2}\min_{\Upsilon^\markov}\|\Upsilon-\Upsilon^\markov\|_2\right)^2\nonumber\\
    &\leq\f{1}{4}\|\Upsilon-\f{\mbb1}{d_\mathsf{S}^{2k+1}}\|_2^2\nonumber\\
    &=\f{1}{4}\left[\tr(\Upsilon^2)-d_\mathsf{S}^{-(2k+1)}\right].
    \label{eq: appendix eta}
\end{align}

The proof of Lemma 3.4 by R. Low~\cite{Low_2009} requires $m$ to be an integer through the multinomial theorem; in the notation of the cited paper, this can be relaxed to be a real number by applying the multinomial theorem for a real power: convergence requires an ordering such that \unexpanded{$|\alpha_t\mbb{E}M_t|>2^{1-n}|\alpha_{t-n}\mbb{E}M_{t-n}|$} for each $n=1,\ldots,{t-1}$ for both the approximate design and Haar expectations.

Let us explicitly write the process $\Upsilon$, defined in Eq.~\eqref{eq: Upsilon} in the main text, as a function of the set of unitaries $\mathfrak{U}:=\{U_i\}_{i=0}^k$, i.e.
\begin{align}
    \Upsilon[\mathfrak{U}]&=\tr_\mathsf{E}[U_k\mc{S}_k\cdots{U_1}\mc{S}_1U_0(\rho\otimes\Psi^{\otimes{k}})U_0^\dg\mc{S}_1U_1^\dg\cdots\mc{S}_kU_k^\dg],
\end{align}
where here implicitly $U_\ell$ stands for $U_\ell\otimes\mbb1_\text{$2k$-ancillas}$ and the maximally entangled states $\Psi$ are taken to be normalized. As the swaps between the system and the $i$\textsuperscript{th} half ancillary system are given by $\mc{S}_i=\sum\mathfrak{S}_{\alpha\beta}\otimes\mbb1\otimes|\beta\rangle\!\langle\alpha|_i\otimes\mbb1$ where $\mathfrak{S}_{\alpha\beta}:=\mbb1_\mathsf{E}\otimes|\alpha\rangle\!\langle\beta|_\mathsf{S}$, this can be written as
\begin{widetext}
\begin{align}
    \Upsilon[\mathfrak{U}] = d_\mathsf{S}^{-k}\sum \tr_\mathsf{E}\left[U_k\FS_{\alpha_k\beta_k} \cdots{U}_1\FS_{\alpha_1\beta_1} U_0 \rho {U_0^\dg}\FS_{\delta_1\gamma_1}{U_1^\dg}\cdots\FS_{\delta_k\gamma_k}{U_k^\dg}\right] \otimes|\beta_1\alpha_1\cdots\beta_k\alpha_k\rangle\!\langle\delta_1\gamma_1\cdots\delta_k\gamma_k|.\label{upsilon def}
\end{align}
\end{widetext}

Now, the standard approach to compute the sum of the moduli of the coefficients of a given polynomial is to evaluate on an argument (here a $d_\mathsf{SE}\times{d}_\mathsf{SE}$ matrix) full of ones (so that all single monomials equal to one) and take each summand to the corresponding modulus. We follow this approach, however, we first notice that the environment part in Eq.~\eqref{upsilon def} is just a product of the environment parts of all unitaries and initial state. To see this, let \unexpanded{$U=\sum{U}^{es}_{e^\prime{s}^\prime}|es\rangle\!\langle{e}^\prime{s}^\prime|$} where \unexpanded{$|e\rangle$ and $|s\rangle$} are $\mathsf{E}$ and $\mathsf{S}$ bases. Unitarity then implies \unexpanded{$\sum\overline{U}_{es}^{ab}U_{\epsilon\sigma}^{ab}=\delta_{e\epsilon}\delta_{s\sigma}$}, where the overline denotes complex conjugate, and so this means that \unexpanded{$\tr_\mathsf{E}[V\mf{S}_{\alpha\beta}U\rho{U}^\dg\mf{S}_{\gamma\delta}V^\dg]=\sum
{V}^{es}_{e^\prime{s}^\prime}\overline{V}^{e\sigma}_{e^\prime\sigma^\prime}{U}^{e^\prime{s}_2}_{b{s}_2^\prime}
\overline{U}^{e^\prime\sigma_2}_{b\sigma_2^\prime}\rho^{br}_{bt}\,\phi(S)$} where \unexpanded{$\phi(S)$} stands for the system $\mathsf{S}$ part; for each $b$ index the rest of the terms are summed over $e$; this generalizes similarly for any number of unitaries. This implies that at most $d_\mathsf{E}$ terms need to be set to one and we can evaluate $\Upsilon$ in a set of matrices $\mc{J}=\{\mbb1_\mathsf{E}\otimes{J}_\mathsf{S},\cdots,\mbb1_\mathsf{E}\otimes{J}_\mathsf{S},J_\mathsf{E}\otimes{J}_\mathsf{S}\}$ with $J$ a matrix with each element equal to one in the respective $\mathsf{E}$ or $\mathsf{S}$ systems: let $\rho= \sum \rho_{ese^\prime{s}^\prime}|es\rangle\!\langle{e^\prime{s}^\prime}|$, then
\begin{widetext}
\begin{align}
    \Upsilon[\mc{J}]&=d_\mathsf{S}^{-k}\sum\rho_{ese^\prime{s}^\prime}\tr[d_\mathsf{E}J_\mathsf{E}|e\rangle\!\langle{e}^\prime|]{J}_\mathsf{S}|\alpha_k\rangle\!\langle\beta_k|\cdots|\alpha_1\rangle\!\langle\beta_1|J_\mathsf{S}|s\rangle\!\langle{s}^\prime|J_\mathsf{S}|\delta_1\rangle\!\langle\gamma_1|\cdots|\delta_k\rangle\!\langle\gamma_k|J_\mathsf{S}\otimes|\beta_1\alpha_1\cdots\beta_k\alpha_k\rangle\!\langle\delta_1\gamma_1\cdots\delta_k\gamma_k|\nonumber\\
    &=\f{d_\mathsf{E}}{d_\mathsf{S}^k}\sum\rho_{ese^\prime{s}^\prime}\,{J}_\mathsf{S}|\alpha_k\rangle\!\langle\beta_k|\cdots|\alpha_1\rangle\!\langle\beta_1|J_\mathsf{S}|s\rangle\!\langle{s}^\prime|J_\mathsf{S}|\delta_1\rangle\!\langle\gamma_1|\cdots|\delta_k\rangle\!\langle\gamma_k|J_\mathsf{S}\otimes|\beta_1\alpha_1\cdots\beta_k\alpha_k\rangle\!\langle\delta_1\gamma_1\cdots\delta_k\gamma_k|,
\end{align}
and hence (we now omit the subindex $\mathsf{S}$ on the $J$ matrices for simplicity),
\begin{align}
    \left(\f{d_\mathsf{S}^k}{d_\mathsf{E}}\right)^2
    \tr[\Upsilon^2(\mc{J})]&=
    %\left(\f{d_\mathsf{E}}{d_\mathsf{S}^k}\right)^2
    \sum\rho_{ese^\prime{s}^\prime}\rho_{\epsilon\sigma\epsilon^\prime\sigma^\prime}\tr[
    {J}|\alpha_k\rangle\!\langle\beta_k|\cdots|\alpha_1\rangle\!\langle\beta_1|J|s\rangle\!\langle{s}^\prime|J|\delta_1\rangle\!\langle\gamma_1|\cdots\nonumber\\
    &\qquad\qquad\qquad\qquad\qquad J|\delta_k\rangle\!\langle\gamma_k|J^2|\gamma_k\rangle\!\langle\delta_k|J\cdots|\gamma_1\rangle\!\langle\delta_1|J|\sigma\rangle\!\langle\sigma^\prime|J|\beta_1\rangle\!\langle\alpha_1|J\cdots|\beta_k\rangle\!\langle\alpha_k|J
    ]\nonumber\\
    &=%\left(\f{d_\mathsf{E}}{d_\mathsf{S}^k}\right)^2
    d_\mathsf{S}^2\sum\rho_{ese^\prime{s}^\prime}\rho_{\epsilon\sigma\epsilon^\prime\sigma^\prime}\tr[J|\alpha_k\rangle\!\langle\beta_k|\cdots|\alpha_1\rangle\!\langle\beta_1|J|s\rangle\!\langle{s}^\prime|J|\delta_1\rangle\!\langle\gamma_1|\cdots\nonumber\\
    &\qquad\qquad\qquad\qquad\qquad \langle\gamma_{k-1}|J|\delta_k\rangle\!\langle\delta_k|J|\gamma_{k-1}\rangle\cdots|\gamma_1\rangle\!\langle\delta_1|J|\sigma\rangle\!\langle\sigma^\prime|J|\beta_1\rangle\!\langle\alpha_1|J\cdots|\beta_k\rangle\!\langle\alpha_k|J]\nonumber\\
    &=%\left(\f{d_\mathsf{E}}{d_\mathsf{S}^k}\right)^2
    d_\mathsf{S}^{2k+1}\sum\rho_{ese^\prime{s}^\prime}\rho_{\epsilon\sigma\epsilon^\prime\sigma^\prime}
    %\nonumber\\&\hspace{0.75in}
    \,\tr[J|\alpha_k\rangle\!\langle\beta_k|\cdots|\alpha_1\rangle\!\langle\beta_1|J|s\rangle\!\langle{s}^\prime|J|\sigma\rangle\!\langle\sigma^\prime|J|\beta_1\rangle\!\langle\alpha_1|J\cdots|\beta_k\rangle\! \langle\alpha_k|J]\nonumber\\
    &=%\left(\f{d_\mathsf{E}}{d_\mathsf{S}^k}\right)^2
    d_\mathsf{S}^{2k+3}\sum\rho_{ese^\prime{s}^\prime}\rho_{\epsilon\sigma\epsilon^\prime\sigma^\prime}
    %\nonumber\\&\hspace{0.75in}
     \langle\beta_k|J|\alpha_{k-1}\rangle\cdots\langle\alpha_2|J|\alpha_1\rangle\!\langle\beta_1|J|s\rangle\!\langle{s}^\prime|J|\sigma\rangle\!\langle\sigma^\prime|J|\beta_1\rangle\!\langle\alpha_1|J\cdots \langle\alpha_{k-1}|J|\beta_k\rangle\nonumber\\
    &=%\left(\f{d_\mathsf{E}}{d_\mathsf{S}^k}\right)^2
    d_\mathsf{S}^{2k+5}\sum\rho_{ese^\prime{s}^\prime}\rho_{\epsilon\sigma\epsilon^\prime\sigma^\prime}
    %\nonumber\\&\hspace{0.75in}
     \langle\beta_{k-1}|J|\alpha_{k-2}\rangle\cdots\langle\alpha_2|J|\alpha_1\rangle\!\langle\beta_1|J|s\rangle\!\langle{s}^\prime|J|\sigma\rangle\!\langle\sigma^\prime|J|\beta_1\rangle\!\langle\alpha_1|J\cdots \langle\alpha_{k-1}|J|\beta_{k-1}\rangle\nonumber\\
    &=%\left(\f{d_\mathsf{E}}{d_\mathsf{S}^k}\right)^2
    d_\mathsf{S}^{2(2k+1)}\sum\rho_{ese^\prime{s}^\prime}\rho_{\epsilon\sigma\epsilon^\prime\sigma^\prime},
\end{align}
\end{widetext}
where to obtain the second line we used the fact that $J^n=d^{n-1}J$ for positive integers $n$, here applied for $n=2$, together with the trace over system $\mathsf{S}$ given by $\sum\langle\gamma_k|\cdot|\gamma_k\rangle$. This is similarly done to get the third line by $\sum|\delta_k\rangle\!\langle\delta_k|=\mbb1_\mathsf{S}$, and taking the trace summing over $|\gamma_{k-1}\rangle$, which can subsequently be done for all $|\gamma_i\rangle$ and $|\delta_i\rangle$. For the fourth line, the cyclicity of the trace was used, followed by an identity taken by summing up over $|\alpha_k\rangle$, using $J^2=dJ$, and taking the trace. This can be done through all remaining steps, giving the last equality.
This, together with Eq.~\eqref{eq: appendix eta}, implies that (now writing simply $i$, $j$ for $\mathsf{SE}$ indices),
\begin{align}
    4\eta&\leq d_\mathsf{E}^2d_\mathsf{S}^{2(k+1)}\left(\sum|\rho_{ij}|\right)^2+\f{1}{d_\mathsf{S}^{2k+1}}
    \nonumber\\&
    \leq{d}_\mathsf{E}^4d_\mathsf{S}^{2(k+2)}\sum|\rho_{ij}|^2+\f{1}{d_\mathsf{S}^{2k+1}}
    \nonumber\\&
    \leq{d}_\mathsf{E}^4d_\mathsf{S}^{2(k+2)}+\f{1}{d_\mathsf{S}^{2k+1}},
\end{align}
where in the second line we used $\|X\|_1^2\leq{d}\|X\|_2^2$ for element-wise norms $\|X\|_p^p=(\sum|x_{ij}|^p)$ and in the third line we used $\|\rho\|_2^2\leq1$.

\begin{widetext}
\subsubsection{Markov's inequality on \texorpdfstring{$\mc{N}_\bdiamond$}{N\_*}}
As $d_\mathsf{S}^{-2k1-1}\mc{N}_\bdiamond\leq\mc{N}_1\leq\sqrt{d_\mathsf{S}^{2k+1}}\mc{N}_2$, also for $0<m\leq{\design}/4$,
\begin{align}
    \mbb{P}_{\design_\epsilon}[\mc{N}_\bdiamond\geq\delta] \leq& \ \mbb{P}_{\design_\epsilon} \left[ \sqrt{d_\mathsf{S}^{3(2k+1)}}\,\mc{N}_2\geq\delta \right]
    %\nonumber\\ &
    =\mbb{P}_{\design_\epsilon} \left[ \mc{N}_2^{\,2m}\geq\f{\delta^{2m}}{d_\mathsf{S}^{3m(2k+1)}} \right]\nonumber\\
    \leq& \f{d_\mathsf{S}^{3m(2k+1)}\,\mbb{E}_{\,\design_\epsilon}\mc{N}_2^{\,2m}}{\delta^{2m}}%\nonumber\\&
    \leq\left(\f{d_\mathsf{S}^{3(2k+1)}}{\delta^2}\right)^m\left[\left(\f{4m}{\mc{C}}\right)^{m}+(2\mc{B})^{2m}+\f{\epsilon}{d_\mathsf{SE}^{\,\design}}\eta^{2m}\right]\nonumber\\
    =&\left(\f{d_\mathsf{S}^{3(2k+1)}}{\delta^2}\right)^m\left\{\left[\f{16m}{(k+1)\,d_\mathsf{SE}}\left(\f{d_\mathsf{S}^{k+1}-1}{d_\mathsf{S}-1}\right)^2\right]^{m}
    %\right.\nonumber\\&\left.
    +(2\mc{B})^{2m}+\f{\epsilon}{16^md_\mathsf{SE}^{\,\design}}\left(d_\mathsf{E}^4d_\mathsf{S}^{2(k+2)}+\f{1}{d_\mathsf{S}^{2k+1}}\right)^{2m}\right\}\label{largedev_design},
\end{align}
where in the third line we used Markov's inequality. This concludes the proof of Theorem~\ref{Result: Large Dev Markov}.
\end{widetext}

\subsection{Convergence towards Markovianity}\label{Appendix: Markov Convergence}
We may first examine the third and penultimate lines leading to Eq.~\eqref{largedev_design} for meaningful bounds $\mbb{P}_{\design_\epsilon}[\mc{N}_\bdiamond\geq\delta]$. The term $d_\mathsf{S}^{3(2k+1)}/\delta^2$ arises from bounding the diamond norm and Markov's inequality; while $\delta$ is arbitrary, the $d_\mathsf{S}^{3(2k+1)}$ could still be relevant when multiplied with $\mbb{E}_{\,\design_\epsilon}\mc{N}_2^{\,2m}$. This latter term will be small provided 1) $\mc{C}$ is large, 2) $\mc{B}$ is small and 3) the unitary design is approximate and high enough.

For 1) and 2), as detailed by Modi et al.~\cite{FigueroaRomero2019almostmarkovian}, we require a fixed $k$ such that $d_\mathsf{E}\gg{d}_\mathsf{S}^{2k+1}$. This implies $\mc{B}\approx0$, so that
\begin{widetext}
\begin{align}
    \mbb{P}_{\design_\epsilon}[\mc{N}_\bdiamond\geq\delta]
    &\lesssim\left(\f{d_\mathsf{S}^{3(2k+1)}}{\delta^2}\right)^m\left\{\left[\f{16m}{(k+1)\,d_\mathsf{SE}}\left(\f{d_\mathsf{S}^{k+1}-1}{d_\mathsf{S}-1}\right)^2\right]^{m} +\f{\epsilon}{16^md_\mathsf{SE}^{\,\design}}\left({d}_\mathsf{E}^4d_\mathsf{S}^{2(k+2)}+\f{1}{d_\mathsf{S}^{2k+1}}\right)^{2m}\right\} \nonumber\\[0.1in]
    &\approx\left\{\left[\f{16m}{\delta^2(k+1)}\f{d_\mathsf{S}^{2(4k+1)}}{d_\mathsf{E}}\right]^{m} +\epsilon\f{d_\mathsf{E}^{8m-\design}d_\mathsf{S}^{m(10k+11)-\design}}{\delta^{2m}16^m}\right\}\label{LargeDev_BigE}.
\end{align}
\end{widetext}
Now, supposing the $\design$-design is exact, i.e. $\epsilon=0$, we require $m\leq\delta^2\f{(k+1)d_\mathsf{E}}{16\,d_\mathsf{S}^{6k}}$, together with $m\leq{\design}/4$. On the other hand if $\epsilon$ is non-zero, we require
\begin{align}
    \epsilon\ll\left[\delta^2\left(\f{2}{d_\mathsf{E}^2d_\mathsf{S}^{(10k+11)/4}}\right)^4\right]^md_\mathsf{E}^{\,\design}d_\mathsf{S}^{\,\design}.
\end{align}

The choice of real $m$ is only restricted by $0<m\leq{\design}/4$, but otherwise is arbitrary. The right-hand side of Eq.~\eqref{largedev_design} is not monotonic in $m$ over all the remaining parameters, so it won't always be optimal for some fixed choice. One may thus optimize the choice of $m$ numerically for each particular case.

\subsection{Efficient circuit unitary designs}\label{Methods: circuit designs}
As mentioned in the main text, we focus on Result 2 of of Winter et al.~\cite{Winter_HamDesign}. To begin with, an efficient approximation for a unitary design on a system composed of $n$-qubits is shown by Winter et al.~\cite{Winter_HamDesign} for a circuit labeled $\mathrm{RDC}(\mc{I}_2)$, where the name stands for Random Diagonal Circuit, and refers to a circuit where $\mc{I}_2=\{I_i\}$ is a set of subsets of qubit labels $I_i\subset\{1,\ldots,n\}$, such that $|I_i|=2$, i.e., at step $i$, $I_i$ picks a pair of qubits, to which a Pauli-$Z$-diagonal gate with three random parameters is applied. This construction can already be seen in the results of Winter et al.~\cite{Winter_2017} as arising from only two types of random diagonal interactions, which can be simplified into a product of $Z$-diagonal ones. 

A particular case which further simplifies things is then denoted by $\mathrm{RDC}_\text{disc}^{(\design)}(\mc{I}_2)$, where the subscript $\text{disc}$ and the superscript $\design$ refer to discrete sets from which the parameters of the diagonal gates will be sampled, and which are determined by a given natural number $\design$. Specifically, all gates in $\mathrm{RDC}_\text{disc}^{(\design)}(\mc{I}_2)$ have the simplified form 
\begin{gather}
(\mathrm{diag}\{1,e^{i\phi_1}\}\otimes\mathrm{diag}\{1,e^{i\phi_2}\}) \ \mathrm{diag}\{1,1,1,e^{i\vartheta}\},
\end{gather}
where $\mathrm{diag}$ denotes Pauli-$Z$ diagonal, and with $\phi_1,\phi_2$ chosen independently from the discrete set $\{2\pi\,m/(\design+1):m\in\{0,\ldots,\design\}\}$ and $\vartheta$ chosen from $\{2\pi\,m/(\lfloor{\design}/2\rfloor+1):m\in\{0,\ldots,\lfloor{\design}/2\rfloor\}\}$. We emphasise that this is still a circuit with 2-qubit diagonal gates with only three random parameters each, and therein lies its simplicity.

Now let $\mathsf{H}_n = \mathsf{H}^{\otimes{n}}$ be $n$ copies of the Hadamard gate, then the main Result 2 by Winter et al.~\cite{Winter_HamDesign} states that for an $n$-qubit system, when $\design$ is of order $\sqrt{n}$, a circuit of the form
\begin{gather}
\mc{W}_\ell:= \left( \mathrm{RDC}_\text{disc}^{(\design)}(\mc{I}_2) \ \mathsf{H}_n \right)^{2\ell} \ \mathrm{RDC}_\text{disc}^{(\design)}(\mc{I}_2),
\label{eq: circuit W}
\end{gather}
yields an $\epsilon$-approximate unitary $\design$-design if \begin{gather}
    \ell\geq{\design} - \log_2(\epsilon)/n,
\end{gather}
up to leading order in $n$ and $\design$.

All the 2-qubit gates in each repetition of $\mc{W}_\ell$, except those in $\mathsf{H}_n$, can be applied simultaneously because they commute~\cite{Nakata2017decouplingrandom, Winter_2017}. Thus, as explained in the main text, if $\mc{W}_\ell$ yields an approximate unitary design, the order of the non-commuting gate depth $\mathfrak{D}$ will coincide with the bound on the order of the number of repetitions $\ell$.

\twocolumngrid

\section*{Data availability}
No datasets were generated or analysed during the current study

\section*{Code availability}
The code used in the analysis of the datasets is available from the corresponding authors on reasonable request.

\bibliography{bib-file}

\section*{Acknowledgments}
PFR is supported by the Monash Graduate Scholarship (MGS) and the Monash International Postgraduate Research Scholarship (MIPRS). KM is supported through Australian Research Council Future Fellowship FT160100073 and Discovery Project DP210100597. KM thanks Andreas Winter for an insightful discussion and especially Austin Gleeson for highlighting the forgetfulness of nature some two decades ago.

\section*{Author Contributions}
KM and FAP defined the research topic and supervised the research. PFR performed the analytical calculations and plot visualizations. All authors contributed to the theoretical aspects, interpretation of results and writing of the article.

\section*{Competing Interests}
The authors declare no competing interests.

\end{document}